\documentclass[12pt]{article}
\usepackage[a4paper,textwidth=16cm]{geometry} % see geometry.pdf on how to lay out the page. There's lots.
\usepackage{palatino}
\title{The limits of cosmology}
\author{Amedeo Balbi\\{\small \it Physics Department, University of Rome `Tor Vergata', Italy}}
\date{} % delete this line to display the current date

\begin{document}

\maketitle
%\tableofcontents
\abstract{What can we know about the universe? I outline a few of the fundamental limitations that are posed to our understanding of the cosmos, such as the existence of horizons, the fact that we occupy a specific place in space and time, the possible presence of dark components, the absence of a reliable physical framework to interpret the behaviour of the very early universe.}
\subsection*{Introduction}
When attempting to discuss what a certain discipline can or cannot know, one should keep in mind, as a cautionary tale, the famous case of philosopher Auguste Comte. Writing in the first half of the nineteenth century, he stated that astronomers would never be able to ascertain the chemical composition of celestial objects. However, only a few decades after Comte's prediction, Kirchhoff founded spectroscopy and managed to identify chemical elements in the atmosphere of the Sun.

Cosmology is arguably one of mankind's boldest enterprises. It tries to scientifically understand the origin, evolution and structure of the universe as a whole. In doing so, it has to rely on a certain set of observational data (what we {\it see} of the cosmos) whose collection cannot be repeated under  different conditions; furthermore, it has to interpret such data according to a set of physical laws whose validity was mostly assessed in laboratories on Earth.  Most cosmology is based on extrapolations of known physics to uncertain territories, and on indirect evidence derived from the behaviour of the part of the universe we can observe. We happen to live in the golden age of cosmology---for the first time in the history of mankind we are able to scientifically describe the overall structure of the universe. However, to some extent, it is surprising that we have managed to make some sense of the universe at all. 

As researchers, most of the time we optimistically (and pragmatically) focus on what is possible (or will be possible) to achieve through our investigation. Sometimes, however, it is useful to adopt a different perspective and take some time to reflect on what is not possible (or will never be possible). In this essay, I will try to outline, without any presumption of completeness, some of the restrictions which are placed on our knowledge of the universe. Some of them are unavoidable, and for example are related to our specific place in the universe and/or to some limiting principle of physics; some are technical and, remembering the fate of Comte's remark, should be looked at with an open mind; finally, some are related to our currently incomplete knowledge of physics---an incompleteness that may or may not be overcome in the future.

\subsection*{Limited horizons}

It is remarkable that one of the most well-known limitations in physics---namely, the finiteness of the speed of light---has become one of the most powerful tools in cosmology. We can only observe the slice of universe that is connected to us by causal signals (technically, our information on the 3+1 dimensions of spacetime is limited to the surface of the 2+1 null time cone  that extends from our given position back into the past). Given the enormous scale of the universe, this also means that we can observe the universe at different epochs (the farthest regions of the universe looking younger than the closest), thus tracking and reconstructing cosmic history. 

The very possibility of doing cosmology is, in a sense, accidental. Were the universe not evolving in time (as in the now discredited steady-state model) we would not be able to trace any evolution by looking at distant regions. Cosmology would be a rather boring collection of unchanging pictures of objects. 
But the universe does change in time. So, a limitation---the finiteness of the speed of light---has turned into a possibility: the possibility of looking at the very different physical states that the universe went through its history.

The finiteness of the speed of light, however, puts another serious limitation to our ability to probe and observe the whole universe. Not only we can just access a limited spacetime surface, but the extent of the observable universe is bounded by an {\em horizon}, the distance that signals have traveled from the big bang. At any given time $t$, the comoving size of the horizon is given by $d_H(t)=c \int_0^{t} {dt'/R(t')}$.
Whether this size is finite or not depends on the functional form of $R$, the scale factor describing the expansion of the universe, which in turn depends on the physical details of the cosmological model. In the standard big bang model, the size of the horizon grows with time, so that at any epoch there are regions of the universe which cannot be accessed by our observations.

The existence of an horizon in the universe might at first seem a philosophical issue rather than a scientific one. After all, the extent of the visible universe today is so large that we might content ourselves with studying it, without worrying to much about the existence of unseen regions outside the horizon.  But cosmology continuously seems to deal with global features of the universe, despite what we can actually try to measure is only their value in our observable patch.

\subsection*{The shape of what?}
One example of such local-to-global extrapolation has to do with the curvature of the universe, which is directly related to its overall content, usually quantified in terms of the parameter $\Omega$. This parameter measures the total density of the universe in units of a critical value $\rho_c$. When $\Omega=1$, the universe has a null curvature, i.e.\ the spatial sections of the spacetime metric can be described by a set of euclidean coordinates. Mathematically, this critical case is the boundary that separates two different classes of geometries, those with positive or negative constant curvature, corresponding to the so-called {\em closed} or {\em open} cosmological models. Such simple classification is so appealing that is often popularized uncritically, in a way that seems to imply that we can actually know the ``shape of the universe''. In fact, what can we know is, at best, the geometry of the universe on scales which are comparable to the size of the horizon. Nothing can be inferred on the global geometry of the universe. 

The most powerful way to determine the curvature on cosmological scales is currently provided by observations of the cosmic microwave background (CMB). This is a thermal radiation filling the universe, the leftover of the early hot phase of the universe. It is also the farthest electromagnetic signal we can receive in the cosmos: the CMB photons have traveled unscattered for almost the entire age of the universe (which is currently estimated at roughly 13.7 billion years), from the epoch of recombination of neutral hydrogen (about 380 thousand years after the big bang). The expansion of the universe has redshifted the CMB photons towards the microwave region of the electromagnetic spectrum. We are then surrounded by a sort of cosmic photosphere, covering the whole sky, situated at a distance comparable to that of the present horizon\footnote{Note that the distance to the CMB photosphere is smaller than the horizon. Since the universe was opaque beyond the photosphere, this further limits the distance we can directly observe---at least using electromagnetic signals.}. Since the CMB photons traveled along the geodesics of the spacetime metric, measured changes in the angular dimension of CMB features of known size can be used to trace the geometry of the universe. One well-known feature that is left imprinted in the CMB pattern is the sound horizon at recombination, a quantity that controls the typical size of CMB anisotropies (i.e.\ the slight variations of the CMB intensity across the sky). By determining the angular size of the sound horizon, CMB observations managed to measure the $\Omega$ parameter for the first time, giving a value that is consistent with a null curvature, $0.99<\Omega<1.05$ \cite{2009ApJS..180..330K}. This is telling us that the observable patch of the universe does not deviate considerably from a euclidean geometry. 

But, ironically, the accurate determination of a value of $\Omega$ so close to unity is limiting the possibility of extrapolating the knowledge of the curvature of the universe outside the horizon. Were the universe very curved, we would have been able, for example, to decide unambiguously that it was closed or open. In the first case, we could have inferred that the universe has a finite size (in the usual two-dimensional analogy, a positively curved universe is visualized as the surface of a sphere); in the latter, we would have concluded that it is infinite and unbounded (just as the surface of an hyperboloid in two dimensions). The fact that the universe is nearly flat, instead, might leave forever unanswered the question of whether it is just marginally closed or marginally open.  Even with future CMB experiment, such as Planck\footnote{http://www.esa.int/Planck}, which will be able to constrain the curvature to within fractions of a percent, the ``true'' geometry of the universe might remain undetermined, unless significant deviations of $\Omega$ from unity are measured.

But things can be even more complicated. If one takes into account the topological features of spacetime (roughly speaking, the way space is connected), then the same geometry can be realized in different ways \cite{2007arXiv0704.3374L}. For example, one might live in a flat or open universe with a `compact' topology. Such universes would be finite. However, if the curvature is very small, or the typical scale of connectedness is larger than the horizon, no observations could ever be able to decide the global geometry of the universe.

\subsection*{Lost memories}

The observed flatness of the universe is considered as an evidence in favour of  {\em inflation}, a period of accelerated expansion in the early universe,  that was first introduced to solve some shortcomings of the classic big bang model \cite{1982PhLB..108..389L}. One of them is the fact that the universe is very homogeneous inside the horizon: a fact that seems puzzling, since, in the classic big bang model, regions which were not previously in causal contact are mysteriously synchronized when they enter the horizon. According to inflation, during some very early epoch, a tiny homogeneous patch expanded enormously on a short time scale, so that it ended up encompassing a much larger region than the horizon size (technically, during inflation the comoving size of the horizon shrunk instead of growing).  This basically washed out most memories of the previous physical state of the universe. After inflation, all universes look roughly the same within the horizon. What we see today in the observable universe, then, might be not typical at all of the ``true universe'' outside the horizon. In principle, for example, curvature might not be constant outside the horizon.

Inflation is a paradigm, rather than a specific model based on well-established physics. While the overall scenario has not been falsified by any cosmological observations so far, and no alternative viable model can currently explain the same features of our universe as well as inflation does (why the universe is nearly flat, why it is so uniform, why there is a scale-invariant primordial spectrum of density perturbations), testing specific models is hard, just because most details are lost when inflation ends. Most proposed tests (such as the level of non Gaussianity in primordial perturbations, or the presence of a background of gravitational waves that might be detected in the polarized component of the CMB) require observations that currently are just nearly at or beyond the limits of reachable accuracy. Whether inflationary models can be actually tested with future observations remains a big unknown of cosmology.

Even worse, one of the most popular scenarios (\cite{1983PhLB..129..177L,1986PhLB..175..395L}), the so-called {\it chaotic inflation}, predicts that our observable universe is just a patch of spacetime which inflated within a larger (possibly infinite) universe; furthermore, different patches in such `global' universe might have inflated (or will inflate) at different times starting from very different initial conditions. This would lead to an enormous set of casually disconnected regions, each one a universe in its respect, with different physical features. Our cosmology would be strictly  limited to the description of our local patch, leaving forever unanswered the question of what the conditions might be outside the horizon. Although this might at first seem as a reasonable compromise, it could turn into a serious setback if the features of our universe were really just a random occurrence extracted from a larger (perhaps infinite) set of possibilities. Any fundamental understanding of why the universe is as it is would be essentially impossible. 

\subsection*{Shaky foundations}

Our cosmological model is based on the extrapolation of known physics at very large scales and at very high energies. Direct observation (based on electromagnetic signals) of the universe is impossible beyond the epoch of reionization, when the CMB was emitted, although we can indirectly infer the status of the universe at earlier times. Using gravitational waves or neutrinos as probes of earlier epochs, although in principle possible, is still technically unfeasible. 

We believe we have a very solid picture (based both on sound, well-tested physics and on reliable observations) as far back as the epoch of primordial nucleosynthesis, when the primordial abundance of chemical elements was frozen. How far back we can safely push our physical models, however, is hard to tell. The standard model of particle physics should work to describe the universe up to energies of roughly 250 Gev, which are thought to be present at $t\sim 10^{-12}$ s after the big bang. At earlier times, physical theories are not well established. In fact, testing  unified models of fundamental interactions---either quantum gravity or string theory---that should be applied to the universe near the Planck epoch (at $t\sim 10^{-43}$ s, when energies are $\sim 10^{19}$ GeV) seems currently hopeless. 

For string theory, the case appears even more desperate since the latest development in the field seem to predict a humongous set of solutions (something as big as $10^{500}$) without any fundamental preference for a particular one. Each solution would result in the evolution of a different universe with different properties. Such so-called {\em landscape} (see \cite{2003dmci.confE..26S}) would certainly be the land of opportunities: everything physically possible would happen somewhere. But the physical scenario producing the landscape would be intrinsically untestable, since no observation we can perform on our universe could impose any constraint to the model. Cosmology would be a descriptive science.

\subsection*{Obscure matters}

What we know of the universe is limited by the amount of information we can collect. Currently, our only channel to infer the status of distant regions of the cosmos is electromagnetic radiation  in its various forms. However, a well known conclusion of mainstream cosmology is that the vast majority of mass and energy in the universe emits little or no electromagnetic radiation, thus escaping a direct observation. The existence of dark matter and dark energy, which constitute roughly 1/3 and 2/3 of the total density of the universe, is only inferred through the gravitational effect they exert on the structure of the spacetime or on clustered visible matter. Dark matter is needed to explain the large-scale distribution of matter in the universe and the observations performed on clumpy structures, such as, for example, in clusters of galaxies. Dark energy is postulated to explain the observed acceleration of the expansion of the universe inferred by the luminosity of distant type Ia supernovae. Furthermore, both dark matter and dark energy are needed to explain why the total density of the universe is so close to the critical value, given that the bounds on the amount of visible matter in the universe are very much lower than that.

Our picture of the universe then relies heavily on the putative contribution from two components whose nature is totally unknown. While there are reasonable hopes that someday dark matter, if in the form of weakly interacting massive particles,  might be directly detected in the laboratory, a direct characterization of dark energy seems far-fetched. What we know or can hope to know on dark energy is expected to come from cosmological observation. This is a somewhat uncomfortable situation, since the interpretation of cosmological observations is based on a physical model that postulates the existence of those very unknown components that is trying to constrain.  

In any case, if we assume that something like dark energy exists in the universe, some of the simple statements about the universe that are usually derived from classical cosmology lose most of their value. 

For example, the existence of dark energy complicates the problem of accurately determining the curvature of the universe. As we mentioned earlier, the geometry of the universe is determined by relating angular and proper sizes through the angular diameter-distance relation. While in classical cosmology this only depends on the total amount of matter in the universe, when dark energy is taken into account a so-called {\em geometrical degeneracy} is introduced. When observing the CMB, for example, this results in the difficulty of distinguishing the effect of curvature from the effect caused by different models of dark energy: for example, in the simplest case, those characterized by different values of the equation of state parameter $w$ which relates the pressure and density of dark energy ($w=p/\rho$) \cite{2003ApJ...588L...5B}. Accurate determination of the curvature of the universe, then, requires previous assumptions on the model of dark energy. On the other hand, to further complicate things, most current constraints on dark energy derived from observations assume that the universe is spatially flat. 

Another indeterminacy that results from postulating the existence of some form of dark energy is the one related to the destiny of the universe. It was classically assumed that the future evolution of the universe was tightly related to its total matter content or, equivalently, to its geometry. An open universe (having a density smaller than the critical value) would expand forever, while a closed one (with a larger than critical density) would eventually recollapse. This is not true anymore if the universe contains dark energy, which drives an accelerated expansion. The future dynamics becomes unrelated to the matter content, since gravity does not act only attractively when dark energy is present.  Actually, as pointed out in \cite{1999GReGr..31.1453K}, if dark energy exists, no set of observations will ever be able to ascertain the ultimate fate of the universe.

\subsection*{Stranded on this island}

We observe the universe from a specific place in space and time. Although we have long known that our position in the universe is not central, we cannot say it is completely typical. We are bound to observe the universe from a place having the right conditions for the emergence of intelligent observers, and this, as far as we know, puts stringent prerequisites on our local environment. Most locations in the universe would be rather unfit to harbour intelligent life. Still, we must assume that the universe is such that its local details can be neglected when studying its overall structure. In fact, we base our understanding of the universe on the so-called {\em cosmological principle}, which essentially states that the universe is the same everywhere (on average) so that it should look the same to any observer at a given time. This postulate is not directly verifiable, since we cannot actually change our position in the universe; but it is a reasonable consequence of the fact that the universe does indeed look the same (on average) to us when we observe different directions in the sky, and that we do not think our position is special or central in any sense. 

The cosmological principle is a crucial ingredient of our vision of the universe and strongly limits the models that we can build from general relativity in order to describe the large-scale structure of spacetime. However, violations of the cosmological principle (namely, the hypothesis that we live near the center of an underdense region of the universe, see e.g.\ \cite{2008PhRvL.101m1302C}) has recently been invoked in order to explain the observed accelerated expansion of the universe without postulating the existence of dark energy. There is hope that future observations can be used to decide between the two alternatives, but it is interesting to see that profound consequences may arise when we relax an assumption that is usually taken for granted.

Typical as it may be, though, our position in the universe puts strong limitations to the observations we can perform. This is especially true when dealing with properties of the universe which are statistical in nature, such as those related to the distribution of primordial density perturbations. For simplicity, let us focus on the example of the pattern of CMB anisotropies. This can be thought of as a random field, since we have no way of predicting in a deterministic sense which CMB intensity will a certain model produce in a given direction in the sky as looked from our position in the universe. What our models predict is a statistical quantity, the angular power spectrum $C_l$, which is an average over an infinite number of hypothetical realizations of the CMB sky (think of it as the average properties of the CMB if it were looked from any possible position in the universe). This theoretical quantity then has to be compared to one specific sample of CMB anisotropy, the one we observe on our sky. But the fact is, we will never be able to perform the averaging operation on real observations, since we only have one sky to observe. In practice, this means that the reconstruction of the $C_l$ from observations will always be imperfect, no matter how accurate our measurements. This so-called {\em cosmic variance} is unavoidable, and limits the accuracy to which we can infer the physical parameters of our cosmological models from observations\footnote{One well-known direct consequence of cosmic variance is the difficulty of using CMB alone to constrain dark energy models.}. This is only worse if we cannot observe the entire sky, but only some finite patch. 

\subsection*{Time is of the essence}

Often neglected is the fact that we observe the universe at a specific epoch in cosmic history. In an evolving universe, this places limitations on what we can infer, since in such a universe different epochs are not equivalent. The cosmological principle does not apply in time. Again, the existence of dark energy poses a conundrum, since the value of its density defines the epoch at which cosmic acceleration ensues. According to current observations, we are apparently observing the universe just in the very special epoch which immediately follows the beginning of acceleration. This is puzzling in at least two ways. First, we have no fundamental motivation for the observed value of dark energy, which is actually at least $10^{60}$ times smaller than estimated from fundamental arguments. The near equivalence between the epoch at which acceleration starts and the present epoch seems utterly coincidental and remains unexplained. One might argue \cite{1998astro.ph.11461B} that the timescale for the appearance of observers is necessarily related to the cosmological environment, so that the universe can only be observed when it is roughly $10^{10}$ years old. One fashionable explanation of the value of dark energy density and of other cosmological coincidences is to postulate the existence of a multitude of unconnected universes (multiverses) where the value of dark energy density (and perhaps other parameters) assume random values (see, e.g.\ \cite{2009PhRvD..80f3510B}). In such scenarios, the observed value of dark energy is justified by anthropic consideration, because it would be one of a few that are compatible with the presence of intelligent observers in the universe. This is motivated either  by the above mentioned chaotic inflation scenario (or other similar inflationary models predicting a huge number of inflating spacetime `bubbles'), or by the above mentioned string theory landscape. Whether such predictions are at all testable in a scientific manner is a highly controversial and widely debated subject. It seems safe to say, however, that if this the way the universe is, our future cosmological theories would have to give up much of their explanatory power.

In any case, a second puzzling aspect of the existence of dark energy is that its very existence is only observable at a particular cosmic epoch. An observer in the far past would have had no way of inferring the presence of such a component from cosmological data, and would have drawn wrong conclusions from the theoretical model. This is an interesting fact to keep in mind, since we might actually be in a similar situation in the present universe, if a component that is currently subdominant should come to dominate the energy budget in the future.  Even more intriguingly, it has been shown that the presence of dark energy would make cosmology as a science basically impossible for an observer in the distant future \cite{2008IJMPD..17..685K}, since the accelerated expansion would erase any trace of the observations which acts as pillars of our cosmological model (the expansion of the universe, the existence of the CMB and the abundance of light elements). It is a glooming thought, but it should help put things in the right perspective when we presume too much of our models. 

\bibliography{bib}

\begin{thebibliography}{10}

\bibitem{2009ApJS..180..330K}
E.~{Komatsu}, J.~{Dunkley}, M.~R. {Nolta}, C.~L. {Bennett}, B.~{Gold},
  G.~{Hinshaw}, N.~{Jarosik}, D.~{Larson}, M.~{Limon}, L.~{Page}, D.~N.
  {Spergel}, M.~{Halpern}, R.~S. {Hill}, A.~{Kogut}, S.~S. {Meyer}, G.~S.
  {Tucker}, J.~L. {Weiland}, E.~{Wollack}, and E.~L. {Wright}.
\newblock {Five-Year Wilkinson Microwave Anisotropy Probe Observations:
  Cosmological Interpretation}.
\newblock {\em The Astrophysical Journal Supplements}, 180:330--376, 2009.

\bibitem{2007arXiv0704.3374L}
J.-P. {Luminet}.
\newblock {Geometry and Topology in Relativistic Cosmology}.
\newblock In {\em New Trends in Geometry, and Its Role in the Natural and Life
  Sciences}, 2007.

\bibitem{1982PhLB..108..389L}
A.~D. {Linde}.
\newblock {A new inflationary universe scenario: A possible solution of the
  horizon, flatness, homogeneity, isotropy and primordial monopole problems}.
\newblock {\em Physics Letters B}, 108:389--393, February 1982.

\bibitem{1983PhLB..129..177L}
A.~D. {Linde}.
\newblock {Chaotic inflation}.
\newblock {\em Physics Letters B}, 129:177--181, September 1983.

\bibitem{1986PhLB..175..395L}
A.~D. {Linde}.
\newblock {Eternally existing self-reproducing chaotic inflanationary
  universe}.
\newblock {\em Physics Letters B}, 175:395--400, August 1986.

\bibitem{2003dmci.confE..26S}
L.~{Susskind}.
\newblock {The Anthropic Landscape of String Theory}.
\newblock In {\em The Davis Meeting On Cosmic Inflation}, March 2003.

\bibitem{2003ApJ...588L...5B}
A.~{Balbi}, C.~{Baccigalupi}, F.~{Perrotta}, S.~{Matarrese}, and N.~{Vittorio}.
\newblock {Probing Dark Energy with the Cosmic Microwave Background: Projected
  Constraints from the Wilkinson Microwave Anisotropy Probe and Planck}.
\newblock {\em The Astrophysical Journal Letters}, 588:L5--L8, May 2003.

\bibitem{1999GReGr..31.1453K}
L.~M. {Krauss} and M.~S. {Turner}.
\newblock {Geometry and Destiny}.
\newblock {\em General Relativity and Gravitation}, 31:1453, October 1999.

\bibitem{2008PhRvL.101m1302C}
T.~{Clifton}, P.~G. {Ferreira}, and K.~{Land}.
\newblock {Living in a Void: Testing the Copernican Principle with Distant
  Supernovae}.
\newblock {\em Physical Review Letters}, 101(13):131302, September 2008.

\bibitem{1998astro.ph.11461B}
J.~D. {Barrow}.
\newblock {Cosmology and the Origin of Life}.
\newblock In {\em Varenna Conference on The Origin of Intelligent Life in the
  Universe}, 1998.

\bibitem{2009PhRvD..80f3510B}
R.~{Bousso}, L.~J. {Hall}, and Y.~{Nomura}.
\newblock {Multiverse understanding of cosmological coincidences}.
\newblock {\em Physical Review D}, 80(6):063510, September 2009.

\bibitem{2008IJMPD..17..685K}
L.~M. {Krauss} and R.~J. {Scherrer}.
\newblock {The Return of a Static Universe and the End of Cosmology}.
\newblock {\em International Journal of Modern Physics D}, 17:685--690, 2008.

\end{thebibliography}
\bibliographystyle{unsrt}

\end{document}